\documentclass[prb,superscriptaddress,showpacs,twocolumn]{revtex4}
\usepackage{graphicx}
\begin{document}

\title{Collective vortex rectification effects in films with a periodic
pinning array}

\date{\today}

\author{Cl\'ecio C. de Souza Silva}
\affiliation{Nanoscale Superconductivity and Magnetism Group,
Laboratory for Solid State Physics and Magnetism, Katholieke
Universiteit Leuven, Celestijnenlaan 200 D, B-3001 Leuven,
Belgium}

\author{J. Van de Vondel}
\affiliation{Nanoscale Superconductivity and Magnetism Group,
Laboratory for Solid State Physics and Magnetism, Katholieke
Universiteit Leuven, Celestijnenlaan 200 D, B-3001 Leuven,
Belgium}

\author{B. Y. Zhu}
\affiliation{Quantum Phenomena Observation Technology Laboratory,
The Institute of Physical and Chemical Research(RIKEN) and
Advanced Research Laboratory Hitachi Ltd. Hatoyama, Saitama
350-0395, Japan}

\author{M. Morelle}
\affiliation{Nanoscale Superconductivity and Magnetism Group,
Laboratory for Solid State Physics and Magnetism, Katholieke
Universiteit Leuven, Celestijnenlaan 200 D, B-3001 Leuven,
Belgium}

\author{V. V. Moshchalkov}
\email{Victor.Moshchalkov@fys.kuleuven.ac.be}
\affiliation{Nanoscale Superconductivity and Magnetism Group,
Laboratory for Solid State Physics and Magnetism, Katholieke
Universiteit Leuven, Celestijnenlaan 200 D, B-3001 Leuven,
Belgium}

\begin{abstract}
The vortex ratchet effect has been studied in Al films patterned
with square arrays of submicron antidots. We have investigated the
transport properties of two sets of samples: (i) asymmetrical
antidots where vortices are driven by an unbiased ac current, and
(ii) symmetrical antidots where in addition to the ac drive a dc
bias was used. For each sample, the rectified (dc) voltage is
measured as a function of drive amplitude and frequency, magnetic
field, and temperature. As unambiguously shown by our data, the
voltage rectification in the asymmetric antidots is induced by the
intrinsic asymmetry in the pinning potential created by the
antidots, whereas the rectification in the symmetric antidots is
induced by the dc bias. In addition, the experiments reveal
interesting collective phenomena in the vortex ratchet effect. At
fields below the first matching field ($H_1$), the dc voltage--ac
drive characteristics present two rectification peaks, which is
interpreted as an interplay between the one-dimensional motion of
weakly pinned incommensurate vortex rows and the two-dimensional
motion of the whole vortex lattice. We also discuss the different
dynamical regimes controlling the motion of interstitial and
trapped vortices at fields slightly higher than $H_1$ and their
implications for the vortex ratchet effect.

\end{abstract}


\maketitle

\section{Introduction}

Ratchets are systems with an asymmetric periodic potential capable
of promoting unidirectional transport of classical or quantum
objects subject to an unbiased fluctuating drive. For that
purpose, the external forcing must be a non-equilibrium one, like
an alternating force or a colored noise, since useful work can not
be extracted from equilibrium fluctuations only. The ratchet
effect was first used to explain biochemical mechanisms such as
the intracellular transport of macromolecules.\cite{Magnasco03}
Since then, parallel to advances in the theoretical
understanding,\cite{Reimann_Rev} several devices have been
proposed and fabricated in order to test the ratchet idea at the
classical and quantum levels and to use the ratchet effect for
particle segregation and motion rectification of electrons and
fluxons.\cite{RatchetsAPA} Most of this work, however, is focused
on ratchet systems consisting of single objects or an assembly of
weakly interacting objects. Much less is known about the
collective rectification of strongly interacting systems via the
ratchet effect and work on this subject carried out so far is
mostly theoretical.

Vortices in superconductors with a periodic array of pinning
centers have proven to be an ideal system in which to study
collective dynamical phenomena of particles moving in a periodic
potential. Several interesting dynamical effects have already been
studied in these systems. For instance, as demonstrated
theoretically\cite{Clecio02} and
experimentally\cite{Alejandro03,Villegas03a,Wordenweber04},
periodic pinning potentials are able to guide vortex motion into
high symmetry directions of the pinning lattice. Another
remarkable dynamical effect in these systems is the presence of
Shapiro steps in the voltage-current characteristics when an rf
current is coupled to the dc drive.\cite{Lieve99} The steps appear
due to phase-locking between the rf signal and the vortex motion
over the periodic pining structure.

Controllable motion of vortices by means of the ratchet effect was
recently demonstrated in experiments performed on superconducting
samples with periodic arrays of asymmetric pinning
centers\cite{Villegas03b,Joris05} and for asymmetric
configurations of symmetrical pinning sites.\cite{Wordenweber04}
These structures are able to break the symmetry of the vortex
pinning potential and promote vortex motion rectification when an
unbiased ac current is applied. In Ref.~\onlinecite{Villegas03b},
for instance, the vortex ratchet effect was demonstrated in a Nb
film where the asymmetrical pinning potential was given by an
array of triangular magnetic dots. In Ref.~\onlinecite{Joris05},
we observed pronounced rectification effects in Al films where the
symmetry of the pinning potential was broken by engineering a
composite configuration of antidot lattices with big and small
antidots placed close to each other.

Despite a great deal of theoretical
efforts\cite{Lee99,Wambaugh99,Olson01,Zhu01,Zhu04} aimed at the
description of the ratchet dynamics of vortices in asymmetrical
pinning potentials, these systems are far from being completely
understood. The vortex motion itself, in the presence of strong
pinning sites, is still a subject of intense study. Differently
from many other systems, vortices are soft objects; their core is
able to deform to better adjust to the available pinning
potential. Due to this property, the usual approximation of
treating vortices as localized overdamped particles may break down
and the resulting vortex ratchet effect can be quite different
from, for instance, the effects known for overdamped Brownian
ratchets. Indeed, as we have recently shown, the dynamics ruling
the voltage rectification in a superconducting film with an array
of asymmetrical antidots can be described by underdamped equations
of motion.\cite{Joris05} This means that this system can be
regarded as an example of {\it inertia ratchet} systems, even
though the true mass of a vortex is negligible.

In the present paper, we investigate in details vortex motion
rectification in Al films with nanoengineered antidot arrays as a
function of ac excitation, magnetic field and temperature. Two
systems are investigated: (a) square arrays of {\it asymmetrical}
antidots excited by a symmetrical ac current, an intrinsic ratchet
system, and (b) a square array of {\it symmetrical} antidots
excited by an ac current coupled to a dc bias, i.e. a
tilted-potential ratchet. For fields below the first matching
field, $H_1=\Phi_0/a_p^2$ (where $\Phi_0=h/2e$ is the flux quantum
and $a_p$ is the antidot lattice constant), the vortex response in
both systems is characterized by hysteresis, which suggests
underdamped motion, and a double rectification peak at some
incommensurate fields. These results are interpreted in terms of a
minimal inertia ratchet model and molecular dynamics simulations
of ``inertial'' vortices. The simulation results indicate that the
double peak observed at incommensurate fields can be explained in
terms of plastic deformation of the vortex lattice. For fields
higher than $H_1$, our data reveal that the motion of interstitial
vortices is essentially overdamped, in contrast with the unusual
underdamped dynamics of the vortices trapped by the antidots. This
result suggests that the underlying inertial effect is connected
to the strong interaction between the vortices and the antidots
and not to the real mass of the vortices.

The paper is organized as follows. In Sec.~\ref{Sec:Exp}, we
describe the sample preparation, the experimental procedure and
the experimental results. In Sec.~\ref{Sec:Theory}, we present the
details of our underdamped model and the molecular dynamics
simulations. The experimental data, as well as a comparison
between modelling and experiment, are discussed in
Sec.~\ref{Sec:Discussion}. Finally, in Sec.~\ref{Sec:Concl}, we
draw our conclusions.

\section{Experimental results}\label{Sec:Exp}

\subsection{Sample preparation and experimental details}

\begin{figure}[b]
\centering
\includegraphics*[width=8.5cm]{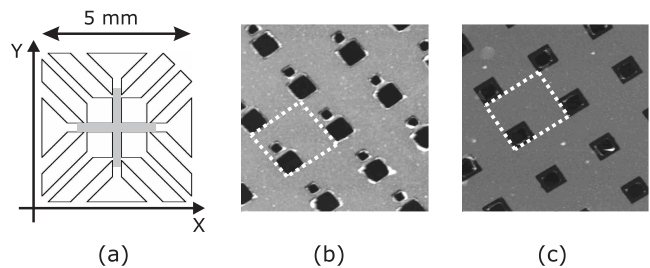}
\caption{\label{sample} Layout of the Al samples. (a) Cross shaped
geometry of the samples. AFM images of sample AAD2 (b) and SAD
(c). The dashed lines indicate the corresponding unit cell.}
\end{figure}

The experiments where performed on three different Al thin films,
thermally evaporated on a SiO$_2$ substrate patterned by electron
beam lithography. For all samples the pattern was designed in a
cross shaped geometry to allow four point measurements in the two
perpendicular directions [see Fig. \ref{sample}.(a)]. The cross
consists of two $300$-$\mu$m-wide strips containing the
nanoengineered array (period $a_p = 1.5$ $\mu$m) of pinning sites.
This gives a value of 0.92 mT  for the first matching field.
The first two samples (AAD1 and AAD2) have the same unit cell
consisting of a small ($300 \times 300$ nm$^2$) and big ($600
\times 600 $ nm$^2$) antidot separated by a thin superconducting
wall (90 nm) and approximate thicknesses (38 and 37 nm). The third
sample (SAD) has a symmetric unit cell consisting of 600 x 600
nm$^2$ square antidots and a thickness of 22 nm.
Figs.~\ref{sample} (b) and (c) show atomic force micrographs (AFM)
of samples AAD2 and SAD, respectively. The dashed lines depict the
unit cell ($1.5\times 1.5$ $\mu$m$^2$) of the respective sample.
The white dots observed in the images are particles of resist
layer which did not come off during the lift off procedure. These
particles are innocuous to our experiment.

Table \ref{tab:table1} shows the superconducting parameters for
all three samples. The superconducting critical temperature $T_c$
was obtained by using a resistance criterion of 10\% of the normal
state resistance in a zero-field measurement. The resistive
transition for all samples was sharp, with a width typically of
$\Delta T_c\sim 3mK$. From the residual resistivity at 4.2 K, the
electron mean free path $l_e$ was found. From this we have
calculated the coherence length $\xi(0)$ and the penetration depth
$\lambda(0)$ using the dirty limit expressions. All samples are
type-II superconductors, i.e., their Ginzburg-Landau parameter is
$\kappa = \lambda(0)/\xi(0) > 1/\sqrt{2}$. Note that, although
samples AAD1 and AAD2 share the same pattern and have similar
thicknesses, their mean free paths were quite different, which is
a result of the different evaporating conditions (base pressure
and growth rate) in which the samples were prepared.

\begin{table}
\caption{\label{tab:table1}Superconducting parameters for the
three samples studied.}
\begin{ruledtabular}
\begin{tabular}{lccr}
&AAD1&AAD2&SAD\\
\hline
Thickness $t$ (nm) & 38 & 37 & 22 \\
Critical temperature $T_c$ (K) & 1.469 & 1.438 & 1.532\\
Mean free path $l_e$ (nm) & 3.0 & 8.1 & 3.4\\
Coherence length $\xi(0)$ (nm) & 59 & 97 & 63\\
Penetration depth $\lambda(0)$ (nm) & 225 & 136 & 210\\
GL parameter $\kappa$ & 3.83 & 1.40 & 3.32\\
\end{tabular}
\end{ruledtabular}
\end{table}

The transport measurements were performed in a cryostat using ac
and/or dc currents applied at the end of the cross legs in the
$x$-direction. The magnetic field was applied perpendicularly to
the $xy$ plane. In all ac measurements, the applied ac current was
a sinusoidal wave, i.e., $I(t) = I_{ac}\sin(2\pi f t)$, where $f$
is the excitation frequency and $I_{ac}$ is its amplitude. The
voltage was acquired in four point mode by an oscilloscope, used
as an analogue-digital converter. The waveform was averaged over
several cycles (typically 200-500 cycles) in order to enhance the
signal-to-noise ratio. The dc voltage was then obtained by
integrating the ``clean'' waveform.

\subsection{Rectification by unbiased asymmetrical antidots}

\begin{figure}[t]
\centering
\includegraphics*[width=8.5cm]{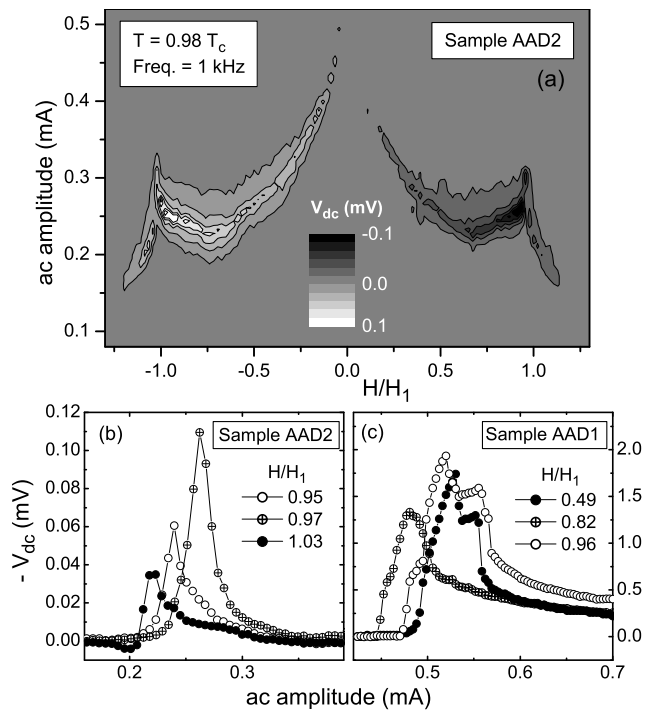}
\caption{\label{fig:VdcAAD} (a) Contour plot of the dc voltage
measured across sample AAD2 as a function of reduced field and ac
amplitude at $T = 0.98T_c$, and frequency of 1 kHz. (b) Detail of
the amplitude dependence of the dc voltage shown in (a) for fields
$H/H_1 = 0.95$, 0.97 and 1.03. (c) Ac-drive amplitude dependence
of the dc voltage measured across sample AAD1 at $T = 0.976T_c$
and $f=1$ kHz for fields $H/H_1 = 0.49$, 0.82 and 0.96.}
\end{figure}

Fig.~\ref{fig:VdcAAD} (a) shows a contour plot of the dc voltage
across the sample ($V_{dc}$) as a function of reduced magnetic
field ($H/H_1$) and current amplitude ($I_{ac}$) at a temperature
$T=0.98T_c$ and excitation frequency $f=1$ kHz. At $H=0$, one can
observe a clear sign inversion of the dc voltage. Note that the
voltage drop due to vortex motion is given by
$V_{dc}=L\cdot(\langle{v}\rangle\times B)$, where $B$ is the flux
density, $\langle{v}\rangle$ is the time averaged vortex velocity,
and $L$ is the distance between the voltage contacts. Thus, the
change of $V_{dc}$ sign as $H$ (and, consequently, $B$) changes
its sign means that vortices and antivortices are rectified to the
same direction. This simple assumption demonstrates unambiguously
the intrinsic rectifying properties of the asymmetrical antidots
configuration. A similar behavior of $V_{dc}(H,I_{ac})$ was also
observed in sample AAD1 and its plot for the positive field side
was presented in Ref.~\onlinecite{Joris05}. Another important
effect demonstrated by Fig.~\ref{fig:VdcAAD} (a) is a clear
enhancement of the rectified voltage near the first matching field
(see also Ref.~\onlinecite{Joris05}, Fig. 2). This can be
explained by the high symmetry of the vortex lattice at this field
range which induces coherent vortex motion, that is, all vortices
contribute to the rectified voltage in a constructive way.

In panels (b) and (c) of Fig.~\ref{fig:VdcAAD}, we show detailed
plots of the $V_{dc}$-$I_{ac}$ characteristics of samples AAD2 and
AAD1, respectively. Similarly to single-object ratchet systems,
these curves are characterized by a rectification window, in which
the rectified voltage increases monotonically, and by the presence
of a tail, in which $V_{dc}$ drops smoothly towards zero. The
rectification window is defined by two critical forces, $F_1$,
which marks the onset of rectification, and $F_2$, for which
rectification is maximum. These forces also determine the
effective asymmetry of the pinning potential, which we define here
as $\alpha = 1 - F_1/F_2$. For some field values (usually, for
fields close to $H_1$ or $H_1/2$) a second rectification peak is
observed, suggesting an interplay between two ratchets (c.f.
Sec.~\ref{Sec:Discussion}), each having their own critical forces
[see Fig.~\ref{fig:VdcAAD}(c)]. However, a clear second peak was
not identified in the measurements on sample AAD2, which is
probably due to a lack of resolution in our measurements in the
narrow rectification window of this sample.

For fields higher than $H_1$, a reversal in the net vortex motion
is observed at lower amplitudes [see the $V_{dc}(I_{ac})$ curve at
$H/H_1=1.03$ in Fig.~\ref{fig:VdcAAD}(b)]. In
Ref.~\onlinecite{Villegas03b}, a similar voltage reversal was
observed at fields above $3H_1$, which was the saturation field of
the sample studied. The authors interpreted this result by arguing
that the vortices trapped by the triangular magnetic dots arrange
themselves in such a way that the interstitial vortices experience
an effective asymmetric potential with opposite asymmetry. In our
experiment, the saturation field was $H_1$. This can be inferred
from the strong drop of the critical current just above this
field, which indicates that weakly pinned interstitial vortices
come into play. Then, in fact, the voltage reversal observed at
fields higher than $H_1$ can also be interpreted as a reversal in
the rectification of interstitial vortices. However, in our case,
the reverse asymmetry is not induced by an asymmetric arrangement
of vortices inside the pins, but rather by the current
distribution (encircling the asymmetric antidot configuration)
created by a single-quantum vortex trapped in the double-antidot
pins.

By comparing the results obtained on samples AAD1 and AAD2, one
can notice the striking difference in the rectified voltage of
both samples. For the similar reduced temperatures, the maximum dc
voltage of AAD2 was several times smaller than that of AAD1. This
can be explained by the fact that the coherence length of AAD2 at
temperatures close to $T_c$ may be bigger than the antidots (for
instance $\xi_{AAD2}(T=0.97T_c)\simeq 560$ nm). In this situation,
the vortices are not able to resolve the details of the antidots
and the effective asymmetry is strongly suppressed. This is very
clear by calculating the asymmetry ratio of both samples at a
relatively low temperature $T=0.976T_c$ (where thermal escape can
be neglected) and reduced field $H/H_1=0.96$, which gives
$\alpha\approx 0.09$ for AAD1 and $\alpha\approx 0.02$ for AAD2.

\begin{figure}[tb]
\centering
\includegraphics*[width=8.5cm]{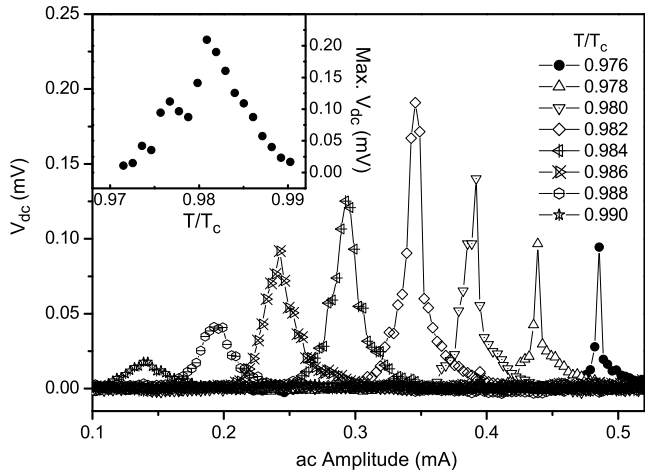}
\caption{\label{fig:Temp-Dep} Dc voltage measured across sample
AAD2 versus ac amplitude for $H/H_1=0.95$ and $f=1$ kHz at
different temperatures, as indicated. Inset: Plot of the maximum
dc voltage at this field as a function of temperature.}
\end{figure}

The coherence length, can also play an important role in the
temperature dependence of the rectified voltage.
Fig.~\ref{fig:Temp-Dep} shows $V_{dc}$-$I_{ac}$ characteristics of
sample AAD2 at several temperatures for the same reduced field
$H/H_1=0.98$.
The maximum rectified voltage seems to increase monotonically
until a temperature $T\simeq 0.982 T_c$, above which it starts to
drop, resulting in a bell-shaped $T$ dependence (see inset). In
Brownian ratchets, a decrease in the dc response as the
temperature increases is indeed expected.\cite{Bartu94} The reason
is that, when the Brownian motion of particles is enhanced by
thermal noise, more energy is needed to move them in a given
direction. However, in the vortex case near the critical
temperature, increasing temperature induces not only vortex
fluctuations but also a strong increase of the vortex size, which
leads to a decrease in the effectiveness of the pinning potential
and its asymmetry. At low temperatures, the reduction of
fluctuations and vortex size should lead to a monotonic increase
in rectification. Nevertheless, when vortices are small enough and
their fluctuations are weak, they are more sensitive to the
background pinning caused by the natural inhomogeneities of the
sample. Due to the disorder induced by the background pinning, the
antidot pinning potential may become less effective at lower
temperatures, thus causing a reduction of the rectified voltage.

\begin{figure}[tb]
\centering
\includegraphics*[width=8.0cm]{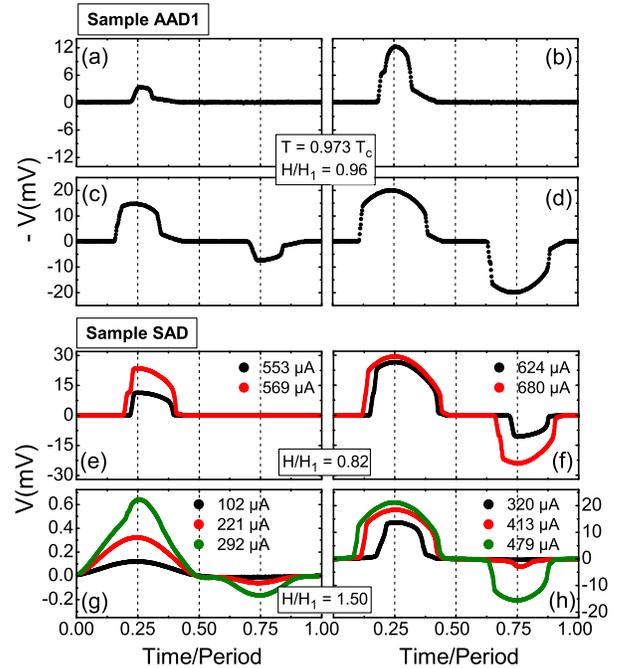}
\caption{\label{fig:outputV} (a)--(d) Time evolution of the
voltage output of sample AAD1 at a temperature $T=0.973T_c$, field
$H=0.96H_1$, and ac amplitudes $I_{ac}=$ 477 (a), 502 (b), 544 (c)
and 704 $\mu$A (d). (e),(f) Voltage output of sample SAD at
$T=0.966T_c$, $H=0.82H_1$, a dc bias $I_{dc}=64$ $\mu$A, and ac
amplitudes: $I_{ac}=$ 553 and 569 $\mu$A (e); $I_{ac}=$ 624 and
680 $\mu$A (f). (g),(h) Voltage output of sample SAD at
$T=0.966T_c$, $H=1.50H_1$, a dc bias $I_{dc}=64$ $\mu$A, and ac
amplitudes: $I_{ac}=$ 102, 221 and 292 $\mu$A (g); and  $I_{ac}=$
320, 413, and 479 $\mu$A (h).}
\end{figure}

In Fig.~\ref{fig:outputV}(a)--(d) we show the voltage output
waveform in one cycle of the ac drive for different drive
amplitudes of the $V_{dc}$-$I_{ac}$ characteristic presented in
Fig.~\ref{fig:VdcAAD}(c) (open circles).
Note that, as discussed above, here negative voltage corresponds
to positive vortex velocity, since the applied field is positive.
Panels (a) and (b) correspond to points in the rectification
window, where the ratchet system behaves as a half-wave rectifier,
whereas (c) and (d) correspond to points in the rectification
tail. At each half-period of the ac drive, the corresponding
half-period of the voltage output in all cases (a-d) is
asymmetric; vortices seem to be depinned at a given force and
repinned at a lower force value, giving rise to hysteretic
behavior. Such a hysteresis was also observed in sample AAD2, but
somewhat smoothed due to the strong effects of fluctuations in
this sample. As we shall see below, this indicates that the system
is ruled by underdamped dynamics.

\subsection{Rectification by dc biased symmetrical antidots}

\begin{figure}[tb]
\centering
\includegraphics*[width=8.5cm]{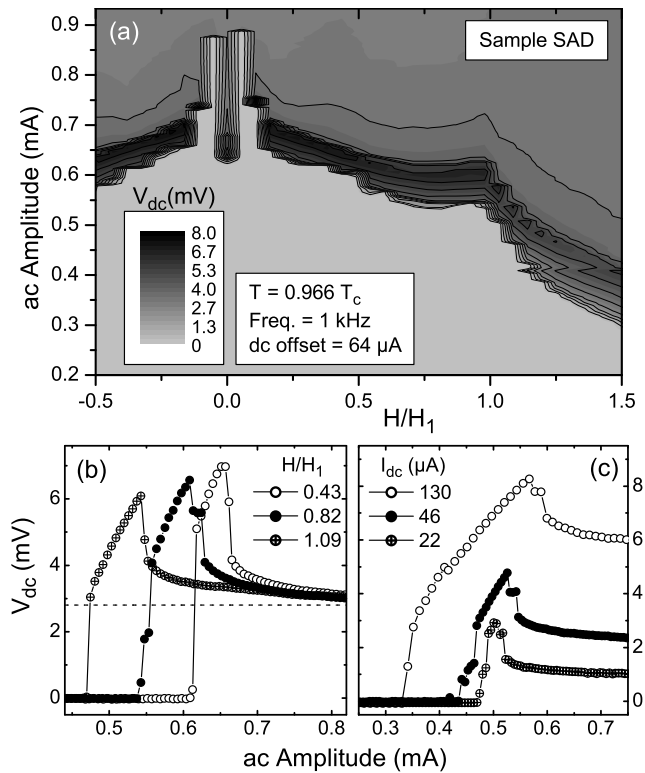}
\caption{\label{fig:VdcSAD} (a) Contour plot of the dc voltage
measured across sample SAD as a function of reduced field and ac
amplitude at $T = 0.966T_c$, frequency of 1 kHz, and a dc bias of
68 $\mu$A. (b) Detail of the amplitude dependence of the dc
voltage shown in (a) for fields $H/H_1=$ 0.43, 0.82, and 1.09. The
dashed line indicates the saturation voltage $V_{\rm
sat}=R_nI_{dc}=2.8$ mV. (c) Ac-drive amplitude dependence of the
dc voltage at $T = 0.973T_c$, $f=1$ kHz, $H/H_1 = 0.93$, and
different tilts, $I_{dc}=$130, 46 and 22 $\mu$A.}
\end{figure}

Since the symmetrical antidot configuration of sample SAD does not
produce rectified vortex motion by itself, we externally induce an
asymmetry in the system by ``tilting'' the pinning potential with
a small dc current bias. In Fig.~\ref{fig:VdcSAD}(a), we show the
contour plot of the dc voltage across sample SAD in the
$H$-$I_{ac}$ plane at a temperature $T=0.966T_c$ for a dc bias of
$I_{dc}=64$ $\mu$A and ac excitations of 1 kHz. [Note that the
steps observed in this plot are just artifacts induced by the low
resolution of the field sweep (0.05 mT).] In this experiment, no
flip of the dc voltage sign is observed. This is indeed expected
because here the polarity of a vortex does matter in determining
its net velocity. For instance, for a positive current bias, i.e.,
in the positive $x$ direction, a vortex with positive polarity
experiences an positive slope in the pinning potential (due to the
Lorentz force pointing to the negative $y$ direction), whereas an
antivortex senses a negative slope. Thus, as the magnetic field
changes its sign, so does the asymmetry of the pinning potential,
in such a way that the cross product $\langle{v}\rangle\times B$
keeps its sign.

In Fig.~\ref{fig:VdcSAD}(b), we present a detailed plot of the
$V_{dc}$-$I_{ac}$ characteristics at fields $H/H_1=$ 0.43, 0.82,
and 1.09, for the experiment of Fig.~\ref{fig:VdcSAD}(a). Here, in
comparison to the $V_{dc}$-$I_{ac}$ characteristics of the
asymmetric antidot samples (Fig.~\ref{fig:VdcAAD}), the important
difference is that the rectification tail approaches
asymptotically a finite voltage value, which depends on the
applied dc tilt, whereas the rectification tail of the
intrinsically asymmetric samples reaches zero voltage at high ac
drives. In the rectification window, however, samples SAD and AAD1
behave in a similar way. For instance, at fields close to $H_1$,
the second rectification peak is also seen in the
$V_{dc}$-$I_{ac}$ curves of sample SAD.

Fig.~\ref{fig:VdcSAD}(c) presents the rectification effect in
sample SAD for different dc bias, $I_{dc} =$ 22, 46, and 130
$\mu$A at $H/H_1=0.93$ and $T/T_c=0.973$. At this field, a double
rectification peak is observed for all dc biases. The distance
between peaks seems to be constant, whereas the rectification
window is wider for higher tilt values. The window is roughly
defined by the critical values $I_c-I_{dc}$ and $I_c+I_{dc}$,
where $I_c=501$ $\mu$A is the critical depinning current obtained
from conventional dc transport measurements. Thus, the system is
given an asymmetry of $\alpha = 2I_{dc}/(I_c+I_{dc})$. For the
tilt $I_{dc}=22$ $\mu$A, this gives $\alpha\approx 0.086$, which
is similar to the intrinsic asymmetry observed in sample AAD1 at
$T=0.973T_c$. Another noteworthy point is that the main
rectification peak decreases relatively to the saturation voltage
for higher tilts. This is consistent with the fact that, in the
limit of the critical tilt ($I_{dc}=I_c$), no rectification peak
should be observed because there the energy barrier for the easy
flow direction disappears completely. In this limiting case,
$V_{dc}(I_{ac})$ is expected to grow monotonically from zero to
the saturation voltage.

The time evolution of the voltage output for different drive
amplitudes of the $V_{dc}$-$I_{ac}$ characteristics shown in
Fig.~\ref{fig:VdcSAD}(b) is presented in
Fig.~\ref{fig:outputV}(e)--(h). We consider two different fields,
$H/H_1 =$ 0.82 (e and f) and 1.50 (g and h), corresponding,
respectively, to filled circles and crossed circles of
Fig.~\ref{fig:VdcSAD}(b). In the $H/H_1 =$ 0.82 case, the
asymmetrical shape of the half-periods of $V(t)$ resembles that
observed in sample AAD1 [shown in panels (a)--(d) of this figure],
that is, here a hysteresis in the depinning-repinning process is
also observed. The main difference is that in the high drive limit
[panel (f)], the amplitude of the positive half-cycle is higher
than the amplitude of the negative one, due to the extrinsic
nature of the asymmetry in sample SAD, whereas for sample AAD1 the
amplitude of both half-cycles are approximately the same.

For fields higher than $H_1$ [Fig.~\ref{fig:outputV} (g) and (h)],
there exists an amplitude range for which the voltage half-cycles
are symmetrical, that is, no hysteresis is observed. Since this
behavior is only present at low ac amplitudes, we can assume that
the symmetrical waveforms correspond to motion of interstitial
vortices. At $H/H_1$ = 1.5, the response is strongly enhanced in
the positive half-cycle above an amplitude $I_{ac}=292$ $\mu$A,
suggesting that at this value the vortices trapped by the antidots
start moving. The voltage waveform corresponding to this amplitude
and above are asymmetrical [panel (h)], resembling the waveforms
observed at fields below $H_1$. A similar evolution of the voltage
waveforms as the ac amplitude is varied was also observed in the
samples with asymmetrical pinning sites for fields $H>H_1$.

\section{Theoretical modelling}\label{Sec:Theory}

\subsection{Inertia ratchet model}

Our experimental data clearly demonstrate a pronounced hysteresis
in the voltage as a transport current is cycled. Typical systems
exhibiting pronounced hysteresis in their transport properties are
driven underdamped particles in periodic potentials. In these
systems, the hysteresis in the depinning-repinning process is a
consequence of the finite inertial mass of the particles, which
has important effects on their transport properties. For instance,
in microscopic boundary lubrication problems, the depinning of the
(inertial) lubricant particles from the periodic potential of a
solid substrate is usually identified as the static friction force
whereas the repinning of the particles is identified as the
dynamical friction force.\cite{Persson93} Other ``inertia''
ratchet systems are asymmetrical underdamped Josephson junction
arrays\cite{Lee03} and dc SQUIDs\cite{SQUIDRatchet}. In these two
examples, the rectified object is the phase of the superconducting
order parameter across a junction and the inertia is given by the
shunt capacitance of the junctions. This problem is formally the
same as that of the mechanical inertia ratchet.

The rectification of inertial particles in a ratchet
potential has been studied theoretically by several groups.\cite{Lee03,%
SQUIDRatchet,Jung96,Borromeo02} The dynamics of inertia ratchets
is more complex then their overdamped counterpart due to the
existence of both regular (for moderate underdamped) and chaotic
(for extremely underdamped) dynamics. In the regular regime, the
amplitude dependent dc response presents a strong enhancement in
the rectification window and a faster decaying tail, as compared
to the characteristics of overdamped particles. Inspired by the
fact that, in our experiment, (i) the rectified voltage
demonstrates hysteretic behavior and (ii) the $V_{dc}$-$I_{ac}$
characteristic generally has pronounced rectification effect in
the rectification window, we propose a description of vortex
dynamics in antidot systems in terms of a simplified underdamped
model, in which vortices (still treated as particles) are given an
apparent mass to account for the observed hysteresis.

We first consider the simpler case of a single vortex in a ratchet
potential, here representing the asymmetric antidot array. The
equation of motion is given by
\begin{equation} \label{eq:motion}
    M\ddot{x} = -\eta\dot{x} - U_p'(x) + A\sin(\omega t),
\end{equation}
where $\eta$ is the viscous drag coefficient, $x$ is the particle
position. The pinning potential $U_p(x)$ is modelled in a
simplified way as the double-sine function:
\begin{equation} \label{eq:Up}
    U_p(x) = -U\left[\sin\left(\frac{2\pi x}{a_p}\right)
    + \frac{\beta}{2}\sin\left(\frac{4\pi x}{a_p}\right)\right]
\end{equation}
The parameter $\beta$ defines the asymmetry of the pinning force
($\alpha=2\beta/(1+\beta)$, for $\beta<1/4$), the critical forces
being given by $F_w=2\pi(1-\beta)U/a_p$, for drive along the
(weak) positive $x$ direction, and $F_s=2\pi(1+\beta) U/a_p$, for
drive along the (strong) negative $x$ direction. Note that, since
here we do not take into account thermal fluctuations, the
critical forces which define the rectification window of this
system are given deterministically by $F_1=F_w$ and $F_2=F_s$.

The main characteristic frequencies setting the limits of the
dynamical regimes of Eq.~(\ref{eq:motion}) are the libration
frequency in the potential minima, which for $\alpha\lesssim 0.2$
is given approximately by $f_p=\sqrt{U/Ma_p^2}$, and the inverse
relaxation time $\gamma=\eta/M$. The quasi-adiabatic regime (i.e.,
close to the zero-frequency adiabatic limit), corresponds to
frequencies much smaller than these two characteristic
frequencies. In the numerical solution of Eq.~(\ref{eq:motion}) we
have chosen excitation frequencies $f<10^{-4}f_p$ and $\gamma\geq
0.1f_p$, corresponding to the quasi-adiabatic regime.

\begin{figure}[tb]
\centering
\includegraphics*[width=8.5cm]{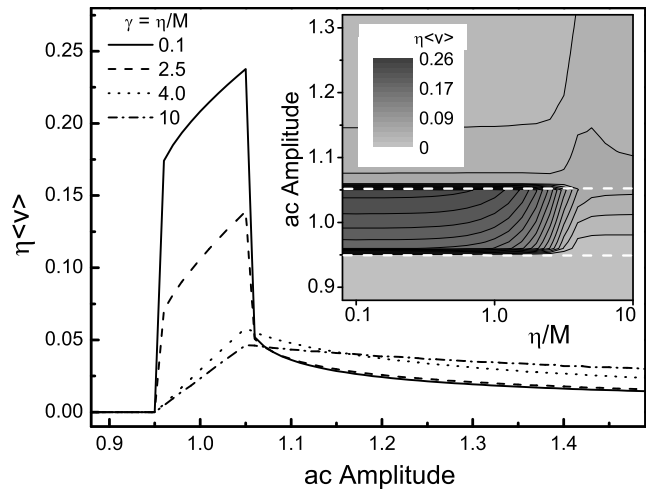}
\caption{\label{fig:IRM} Time averaged particle velocity as a
function of ac drive amplitude calculated in the single-particle
inertia ratchet model [Eq.~\ref{eq:motion}] for the
viscosity-to-mass ratios $\eta/M=0.1$, 2.5, 4 and 10 (in units of
$f_p$). Inset: Contour plot of average velocity as a function of
amplitude and $\eta/M$; the dashed lines define the rectification
window in this case ($\alpha=0.095$).}
\end{figure}

In Fig.~\ref{fig:IRM},
we show the net dc velocity versus ac drive amplitude ($\langle v
\rangle$-$F_{ac}$ characteristics for different values of
$\gamma$. Forces are in units of $2\pi U/a_p$ and frequencies (and
$\gamma$) in units of $f_p$. For these runs, we have chosen the
asymmetry factor $\alpha = 0.095$, which is close to the typical
values of $I_1/I_2$ observed in sample AAD1. The inset shows a
detailed diagram of $\langle v \rangle$ as a function of $A$ and
$M$. Notice the crossover between the spread rectification in the
extreme overdamped limit and the sharp rectification in the
extreme underdamped regime. For $\eta/M\gtrsim 4f_p$, the curves
are essentially overdamped, following the behavior of the
adiabatic solution of a deterministic overdamped
ratchet.\cite{Joris05} For $\eta/M\lesssim 2f_p$, the curves are
much more pronounced in the rectification window and have a faster
decaying tail. We shall compare these results with the
experimental data in Sec.~\ref{Sec:Discussion}.

\subsection{Molecular dynamics simulations}

Now, we consider the 2D problem of a lattice of ``inertial''
vortices interacting with a 2D array of asymmetric pinning sites,
which is modelled as two interpenetrating square sublattices of
strong and weak attractive potentials. The sublattices have
spacing $a_p$ (=1 here) and are shifted with respect to each other
by a distance $d = 0.2 a_p$. All pinning centers are modelled by
Gaussian potential wells with a decay length $R_p$. Hence, the
pinning force exerted by the $k$-th pin of one sublattice on the
$i$-th vortex reads
\begin{equation} \label{Eq:Fp2D}
{\bf F}_p^{s,w}({\bf r}_i) =  - F_{s,w} \,~\frac{{\bf r}_i - {\bf
R}_k^{s,w}}{R_p} \exp \left( - \left\vert \frac{{\bf r}_i - {\bf
R}_k^{s,w}}{R_p} \right\vert^2 \right),
\end{equation}
where ${\bf r}_i$ represents the position of the $i$-th vortex and
${\bf R}_k^{s,w}$ is the location of the $k$-th pin of the weak
($w$) and the strong ($s$) sublattice, respectively. The intensity
of the individual pinning force is denoted by $F_{s,w}$. We define
the ratio between the weak and strong pinning forces as $\epsilon
= F_w / F_s$. In our simulation, forces (per unit length) are
taken in units of $F_0 = \Phi_0^2 /8\pi^2\Lambda^3$, with
$\Lambda$ the effective superconducting penetration depth.
Vortices are considered to interact logarithmically, the
corresponding repulsive vortex-vortex force being given by
$$ {\bf F}_{vv}({\bf r}_i) = F_{vv0} \, \sum_{j \neq i}^{N_v}
{\Lambda \hat {\bf r}_{ij}} / {\vert {\bf r}_i - {\bf r}_j \vert }
\; ,$$
where $\hat {\bf r}_{ij} = ({\bf r}_i -{\bf r}_j) / \vert {\bf
r}_i - {\bf r}_j \vert$. $F_{vv0}$ denotes the strength of the
vortex-vortex interaction. The underdamped equation of motion for
vortex $i$ is given by
\begin{equation}\label{Eq:Motion2D}
M {\ddot{\bf r}}_i  = {\bf F}_L(t) + {\bf F}_{vv}({\bf r}_i)
 + {\bf F}_p({\bf r}_i) - \eta\dot{\bf r}_i \; .
\end{equation}
The ac driving Lorentz force is ${\bf F}_L(t) = A \sin(\omega t)$,
where $\omega = 2 \pi / P$ and $P$ is the period of the ac drive.
$\eta$ is the viscosity coefficient and $M$ is the effective mass
of an individual vortex. This equation is integrated by molecular
dynamics (MD) simulations. Parameters used in the simulations are
$R_p = 0.13 a_0$, $\Lambda = 0.26 a_0$. $F_{vv0} = 0.06$,
$F_{p0}^s = 0.5$ and $\epsilon = 0.925$, time step $\tau_0 =
1/150$. All the frequencies of the ac sine-wave drive are smaller
than $10^{-6}/\tau_0$, i.e., into the quasi-adiabatic regime. For
the mass and viscosity we have chosen, respectively, $M=1$ and
$\eta=1$, which, in terms of $f_p = \sqrt{U_p/Ma_p^2}$ (here $U_p
= F_s R_p^2 /2$ is the amplitude of the individual strong pinning
potential), leads to a viscosity-to-mass ratio $\eta/M=2.44f_p$.
The MD simulation results will be discussed in the next section.

\section{Discussion}\label{Sec:Discussion}

\begin{figure}[bt]
\centering
\includegraphics*[width=8cm]{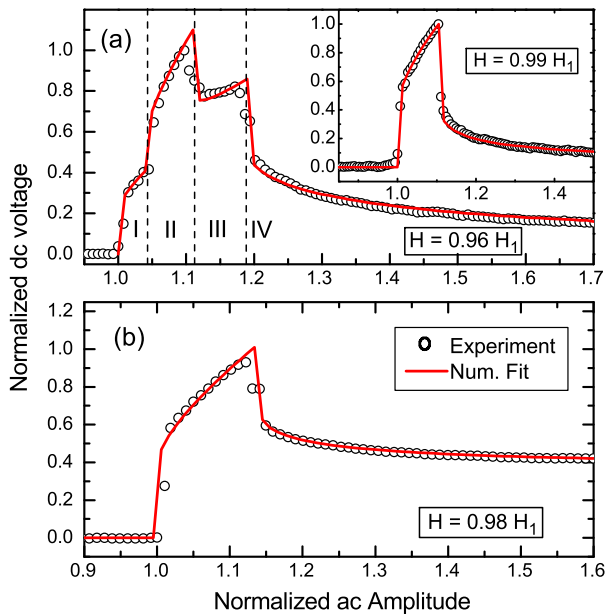}
\caption{\label{fig:InertiaFit}(a) Main frame: normalized dc
voltage versus current amplitude (circles) of sample AAD1 at
$T=0.973T_c$ and $H=0.96H_1$ and numerical fit obtained from the
double ratchet model described in Sec~\ref{Sec:Discussion}(see
also Sec.~\ref{Sec:Theory}) with $\gamma=1.9$ and $\chi=0.5$. The
dashed vertical lines define different dynamical regimes of the
system (see text). Inset: same as the main frame but with
$H=0.99H_1$. The fitting (line) was obtained integrating
Eq.~\ref{eq:motion} with $\gamma=1.8$. (b) Normalized dc voltage
versus current amplitude (circles) of sample SAD at $T=0.966T_c$
and $H=0.98H_1$ using an excitation frequency of 1 kHz and a dc
bias of 46 $\mu$A (which gives a reduced tilt. The fitting (line)
was obtained using $\gamma=2.1$.}
\end{figure}

In Fig.~\ref{fig:InertiaFit}, we present a fitting of our inertia
ratchet model to the experimental data of samples AAD1 and SAD.
The data is plotted in normalized units. The normalized critical
forces used in the model were the same observed in the respective
measurement. In this way, the only adjustable parameter is the
viscosity-to-mass ratio, $\gamma=\eta/M$, which is given in units
of the libration frequency $f_p$. For comparison, we have shown in
Fig.~\ref{fig:IRM} the $V_{dc}$-$I_{ac}$ calculated close to the
overdamped limit $\gamma=10$. As it is clearly seen from this
graph, the usual overdamped model fails to describe the ratchet
effect of vortices in an antidot lattice, more specifically in our
Al samples AAD1 and SAD. On the other hand, the underdamped model
with $\gamma$ values close to the underdamped limit provides very
good fitting to our experimental data. For the measurement of
sample AAD1 at $T=0.973T_c$ and $H/H_1=0.99$, for example, we used
$\alpha=0.095$ (which gives approximately the same normalized
critical forces observed in this experiment) and the best fit was
obtained with $\gamma=1.8$ [see Fig.~\ref{fig:InertiaFit}
(a)--inset], which is close to the underdamped limit. For the
sample SAD at $H/H_1=0.98$ and $T=0.966T_c$, we used $\alpha=0$
(i.e. a symmetrical sinusoidal potential) and added to the driving
force a dc component $F_{dc}=0.066F_p$ (where $F_p=2\pi U/a_p$ is
the critical force of the pinning potential), in agreement with
the $I_{dc}/I_p$ ratio observed in the experiment. In this case
the best fit was obtained with $\gamma=2.1$
[Fig.~\ref{fig:InertiaFit} (b)].

The success of an underdamped model to describe the voltage
induced by vortex motion is not a trivial result. Vortex dynamics
is usually treated in terms of overdamped particle models because
it is a well established fact that in most samples the vortex mass
is negligible compared to the damping coefficient.\cite{mass}
Nevertheless, these models do not consider the effect of vortex
deformation that can lead to an irreversibility in the
depinning-repinning process and, consequently, to an apparent
inertial effect. In recent Ginzburg-Landau simulations of a
superconductor with an antidot lattice, it was shown that the
process of depinning of a vortex from an antidot can be
characterized by strong current-induced elongation of the vortex
core cross-section, which results in a hysteresis in the pinning
mechanism.\cite{Priour03} This effect seems to mimic the behavior
of an underdamped particle in a periodic potential and can be
responsible for the hysteresis observed in our experiments.

\subsection{Vortex rectification for $H<H_1$}

In our experiment, the $V_{dc}$-$I_{ac}$ curves for some fields
below $H_1$ are characterized by a double rectification peak. As
we pointed out in our previous publication, the second
rectification peak can be explained in terms of plastic
deformation of the vortex lattice.\cite{Joris05} In other words,
the existence of a plastic dynamical phase with a lower critical
current gives rise to the appearance of a second ratchet system,
which competes with the main ratchet, given by the motion of the
whole vortex lattice. The consequences of the existence of two
ratchets with different critical forces can be heuristically
illustrated by our one particle underdamped model. Let us consider
two independent inertia ratchets. The first one represents the
whole vortex lattice, with critical forces $F_{w}$ and $F_{s}$.
The second ratchet represents the fraction $\chi$ of the vortices
that deform plastically and has critical forces $F'_{w}\leq F_{w}$
and $F'_{s}\leq F_{s}$. In this way, the total dynamics is simply
given by the solution $v(t)$ of the first ratchet for $F(t)>F_{w}$
and $F(t)<-F_{s}$, and, otherwise, by the solution $v'(t)$ of the
second ratchet, normalized by the factor $\chi$. This simple model
provides an excellent fitting to the experimental data for
$H<H_1$. For instance, to fit the data of the experiment performed
in sample AAD1 at $H=0.96H_1$ and $T/T_c=0.973$, we solved
Eq.~(\ref{eq:motion}) for both ratchets independently, using for
the critical forces the normalized values obtained experimentally,
and adjusted the parameters $\gamma$ and $\chi$. In this case, we
obtained $\gamma=1.9$ and $\chi=0.5$ [see
Fig.~\ref{fig:InertiaFit} (a)].

\begin{figure*}[tb]
    \centering
    \begin{minipage}[b]{.6\textwidth}
        \centering
        \includegraphics[width=1.0\textwidth]{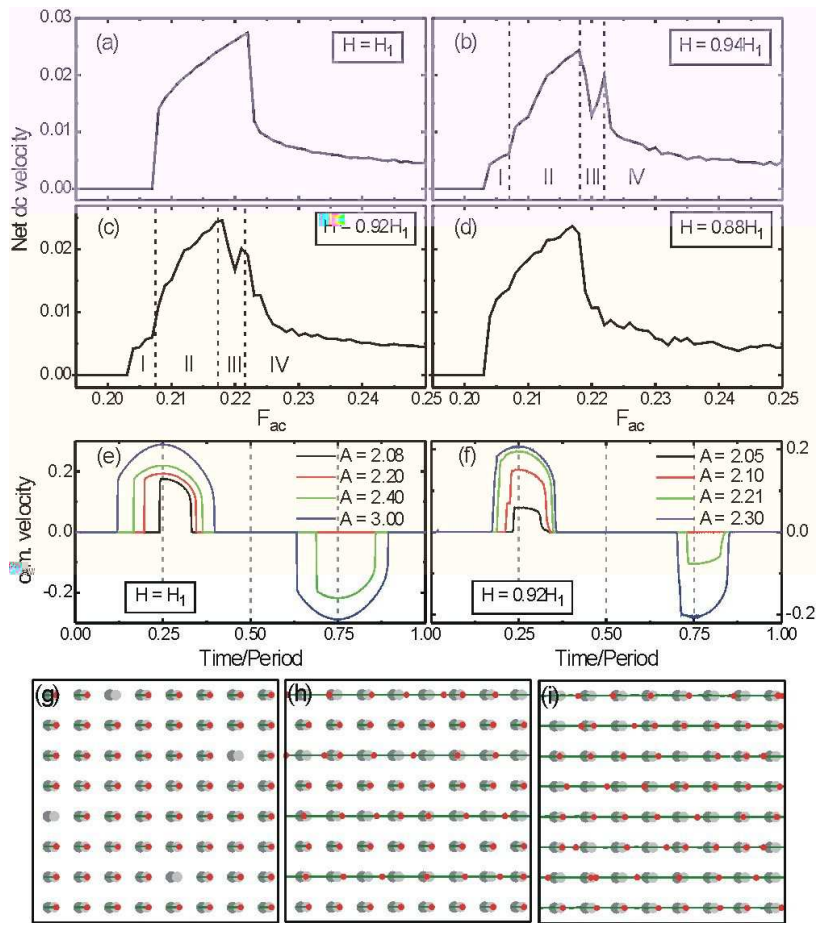}
    \end{minipage}
    \hspace{.05\textwidth}
    \begin{minipage}[b]{.325\textwidth}
        \centering
        \caption{\label{Fig:MD} Molecular dynamics simulations
        results for $\eta=1$ and $M=1$. Panels (a)-(d): net dc velocity as
        a function of the ac driving force amplitude in normalized
        units (see text) at reduced fields $H/H_1=1$ (a), 0.94 (b),
        0.92 (c), and 0.88 (d). The dashed lines in (c) and (d) define
        different dynamical regimes of the system (see text).
        Panels (e) and (f): time evolution
        of the c.m. velocity at $H/H_1=1$ (e) and 0.92 (f). Panels
        (g)-(i): vortex trajectories (lines) for $H/H_1=0.94$ at drive
        amplitudes 0.2 (g), where vortices are still pinned, 0.205
        (h), where part of the vortices move as 1D channels, and
        0.21 (i), where all vortices move. Vortices are represented by
        the small dots and the double pinning sites are represented by
        the double discs.}
    \end{minipage}
\end{figure*}

To get a better insight on how such plastic dynamics can lead to a
double peak in the $V_{dc}$-$I_{ac}$ characteristic, we performed
MD simulations of ``inertial'' vortices for fields $H\leq H_1$.
The parameters used in the simulations are given in
Sec.~\ref{Sec:Theory}-B. Fig.~\ref{Fig:MD} (a)-(d),
show the simulated net velocity of vortices $\langle v\rangle$ as
a function of drive amplitude $F_{ac}$ for reduced fields $H/H_1$
= 1, 0.94, 0.92, and 0.88. The simulations reproduce quite well
the double rectification peak observed experimentally. Panels (e)
and (f) show the time dependence of the center-of-mass (c.m.)
velocity of the vortex lattice at vortex densities $H/H_1=1$ and
0.92, respectively. As in our experiments, here the waveforms of
vortex motion during each half-period are asymmetrical, i.e.,
present hysteretical behavior, which is a result of the finite
apparent mass of the vortices. In particular, the simulated
waveforms at $H/H_1=0.92$ [panel (f)] evolve similarly to the
experimental data of sample AAD1 presented in
Fig.~\ref{fig:outputV} (a)-(d). In both, experiment and
simulations, four different ratchet phases (I-IV) are clearly
identified [these phases are defined by vertical dashed lines in
the $V_{dc}$-$I_{ac}$ characteristic of Fig.~\ref{fig:InertiaFit}
(a) and in the $\langle v\rangle$-$F_{ac}$ characteristics of
Fig.~\ref{Fig:MD} (b) and (c)]. In phase I, only a fraction of
vortices move in the easy direction of the pinning potential. No
motion to the hard direction is observed. In phase II, the whole
vortex lattice is half-wave rectified, i.e., all vortices move to
the easy direction and no vortices move to the hard one. In phase
III, we have motion of all vortices in the easy direction and
motion of a fraction of them to the hard direction. The averaged
rectified voltage drops abruptly at the onset of motion to the
hard direction but increases again at slightly higher amplitudes,
since the main ratchet (the whole vortex lattice) is still in its
rectification window, giving rise to the second rectification
peak. Finally, in the phase IV, all vortices move back and forth,
but in an asymmetric fashion, resulting in the long tail observed
in the experiments and simulations.

To illustrate the interplay between the plastic dynamics and
motion of the whole vortex lattice, we plotted in
Fig.~\ref{Fig:MD} (g)--(i) the vortex trajectories for the
simulation at $H/H_1=0.94$ and a few drive amplitudes. At
$F_{ac}=0.2$ (g), all vortices are still pinned (just oscillating
inside the pinning centers). Note that, since the number of
vortices does not match the number of pinning sites, there is a
distribution of vacant pins, that is to say, of discommensurations
in the vortex lattice. In this case, and also for $H/H_1=$ 0.92,
there are some vortex rows, with a denser distribution of
discommensurations, which depin more easily. Thus, at
$F_{ac}=0.205$ (h), only the weakly pinned vortex rows are dragged
away, giving rise to a one-dimensional channelling dynamics
characterized by a strong shear of the vortex lattice. Finally, at
a higher drive, $F_{ac}=0.210$ (i), the potential barriers for the
strongly pinned vortex rows are also overcome and all vortices
move to the easy direction of the pinning potential. At lower
fields, such as $H/H_1=0.88$ shown in panel (d) of this figure,
the distribution of vacancies is approximately homogeneous in such
a way that all vortex rows has approximately the same depinning
force. Thus, no channelling dynamics is observed and the $\langle
v\rangle$-$F_{ac}$ characteristic at this field presents a single
rectification peak.

\subsection{Vortex rectification for $H>H_1$}

At fields higher than $H_1$, the interpretation of vortex motion
in terms of underdamped dynamics may break down because of the
presence of interstitial vortices. Since they do not interact
directly with the antidots, their dynamics is expected to be
similar to that of vortices in a plain film, that is, essentially
overdamped. This was indeed suggested by the amplitude dependence
of the voltage waveforms $V(t)$ [see Fig.~\ref{fig:outputV} (g)
and (h)]: whereas for high amplitudes the signal in a half-cycle
is asymmetric, at lower amplitudes, the signal may be perfectly
symmetric, that is, completely reversible. Since apparently
hysteresis only come out when the vortices trapped in the antidots
are set into motion, one may then conclude that the apparent
inertia comes from the interaction between vortices and antidots
rather than from the real vortex mass.

\section{Conclusion}\label{Sec:Concl}

In conclusion, we have studied the vortex ratchet effect in
superconducting films with nanoengineered arrays of asymmetric and
symmetric antidots. In the asymmetric antidot sample, we observed
rectification of an unbiased ac current due to the intrinsic
asymmetry of the system, thus demonstrating that this is an
intrinsic vortex ratchet system. In the symmetric antidot sample,
the rectification effect was due to the dc tilt of the periodic
(symmetric) potential induced by the applied dc bias. The vortex
rectification is strongly dependent on temperature and magnetic
field. The temperature dependence indicates that the coherence
length plays an important role in determining the effective
asymmetry produced by the asymmetric antidot array; when the
vortex size becomes comparable to the big antidot size, the
rectification is strongly suppressed. 

The field dependence of the rectification effects found in our
experiments provides the following observations indicating
collective behavior in the ratchet dynamics: (i) rectification is
enhanced near the first matching field, evidencing the importance
of vortex-vortex interactions in the ratchet effect (at the first
matching field, the vortex-vortex interactions cancel out by
symmetry); (ii) at fields above $H_1$, we observed a sign reversal
in the ratchet effect at low amplitudes, which indicates that
weakly pinned interstitial vortices are rectified to the opposite
direction due to the inverse asymmetry produced by the current
distribution of the vortices trapped in the antidots; and (iii) at
some fields below $H_1$, a double rectification peak was observed,
which can be explained by the competition between rectification of
(weakly pinned) incommensurate vortex rows and the rectification
of the whole vortex lattice.

In addition, our data revealed that the ratchet effect for fields
lower than $H_1$ is dominated by underdamped dynamics. These
results were corroborated by molecular dynamics simulations of an
underdamped vortex ratchet model. We argued that the inertial
effect actually comes from vortex deformation due to the strong
interaction with antidots, rather than from a finite vortex mass.
Indeed, at fields higher than $H_1$, we observe that the dynamics
of interstitial vortices is essentially overdamped, since these
vortices are not pinned directly by the antidots.

\section*{Acknowledgement}

We thank Alejandro Silhanek for useful discussions. This work was
supported by the K.U.Leuven Research Fund GOA/2004/02 and FWO
programs. C.C.S.S. is supported by CNPq, an Agency of the
Brazilian Government. M.M. acknowledges support from the Institute
for the Promotion of Innovation through Science and Technology in
Flanders (IWT-Vlaanderen).

\end{document}